\documentclass[prl,showpacs,twocolumn,amsmath]{revtex4}
\usepackage{amssymb}
\usepackage{amsmath}
\usepackage{epsfig}
\usepackage{amsfonts}
\usepackage{color}
\usepackage{float}
\usepackage{latexsym}
\usepackage{mathrsfs}
\usepackage{feynmf}
\def\QED{\leavevmode\unskip\penalty9999 \hbox{}\nobreak\hfill
     \quad\hbox{\leavevmode  \hbox to.77778em{%
               \hfil\vrule   \vbox to.675em%
               {\hrule width.6em\vfil\hrule}\vrule\hfil}}
     \par\vskip3pt}
\def\qed{\leavevmode\unskip\penalty9999 \hbox{}\nobreak\hfill
     \quad\hbox{\leavevmode  \hbox to.77778em{%
               \hfil\vrule   \vbox to.675em%
               {\hrule width.6em\vfil\hrule}\vrule\hfil}}
\par\vskip3pt}
\def\ibb #1{\leavevmode\hbox{\kern.3em\vrule
     height 1.5ex depth -.1ex width .4pt\kern-.3em\rm#1}}

\newcommand{\be}{\begin{equation}}
\newcommand{\ee}{\end{equation}}
\newcommand{\ba}{\begin{array}}
\newcommand{\ea}{\end{array}}
\newcommand{\bqa}{\begin{eqnarray}}
\newcommand{\eqa}{\end{eqnarray}}

\newcommand{\tr}{\mbox{Tr}}

\input amssym.def
\begin{document}
\title{Witness to detect quantum correlation of bipartite states in arbitrary dimension}

\author {Zhi-Hao Ma$^{1}$, Zhi-Hua Chen$^{2}$, Jing-Ling Chen$^{3}$}
\affiliation { Department of Mathematics, Shanghai
Jiaotong University, Shanghai, 200240, P. R. China }

\affiliation {Department of Science, Zhijiang college, Zhejiang
University of technology, Hangzhou, 310024, P.R.China}

\affiliation {Theoretical Physics
Division, Chern Institute of Mathematics, Nankai University,
Tianjin, 300071, P.R.China}

\begin{abstract}
In this work we introduce a nonlinear witness that is a sufficient
condition for detecting the vanishment of quantum correlation of
bipartite states. Our result directly generalizes the result of [J.
Maziero, R. M. Serra, arXiv:1012.3075] to arbitrary dimension based
on the Bloch representation of density matrices.
\end{abstract}


\maketitle

\section{Introduction}

Entanglement is an essential resource in almost all quantum
computing and informational processing tasks
\cite{Horodecki09,Guhne09}. However, there are quantum correlations
beyond entanglement, i.e., entanglement is not necessarily needed to
illustrate the non-localities in a quantum system. It has been shown
that the quantum correlation, even without entanglement, can lead to
the speedup of quantum computing\cite{3}. Furthermore, it is more
robust than entanglement in resisting environment-induced
decoherence, which makes the quantum computation based  on quantum
correlation more robust than those based on the
entanglement\cite{4,4add1,4add2}.

If a bipartite quantum state is in a product state,
$\rho=\rho_A\otimes\rho_B$, with $\rho_A$ ($\rho_B$) being the
reduced density matrix of subsystem A (B), the state has no quantum
correlation. However, a state with zero quantum correlation is not
always a product state.

Detecting whether the quantum correlation of a
bipartite state, quantified by Ollivier and Zurek's quantum
discord\cite{5}, is zero or not is fundamentally important, e.g., it has been proven that zero quantum
discord between a quantum system and its environment is necessary
and sufficient for describing the evolution of the system through a
completely positive map\cite{12,12add1}. In addition, a quantum
state can be locally broadcasted if and only if it has zero quantum
discord \cite{13,13add1}.

In fact, a system is classically
correlated only if its state can be written as
\begin{equation}
\sum_{ij}p_{ij}|a_{i}\rangle \langle a_{i}|\otimes |b_{j}\rangle \langle b_{j}|,  \label{CC}
\end{equation}
with $\{|a_{i}\rangle \}$ and $\{|b_{j}\rangle \}$ forming orthonormal basis
for the two subsystems and $\{p_{ij}\}$ being a probability distribution.

In this work, we will try to solve the problem of detect whether a state is classical or not.

\section{main result}

It is well known that every $N\times N$ density matrix can be
represented by the $(N^2-1)$-dimensional Bloch vector as:
$\rho(\textbf{u})=\frac{1}{N}(I+\sqrt{\frac{N(N-1)}{2}}\overrightarrow{\lambda}.\textbf{u})$,
but the converse is not true, i.e., not all operator of the form
$\frac{1}{N}(I+\sqrt{\frac{N(N-1)}{2}}\overrightarrow{\lambda}.\textbf{u})$
is a density matrix, where $\textbf{u}$ is an arbitrary
$(N^2-1)$-dimensional Bloch vector. Note that a density matrix must
satisfy three conditions: (a). Trace unity,
$\tr(\rho(\textbf{u}))=1$. (b). Hermitian,
$\rho(\textbf{u})^{+}=\rho(\textbf{u})$; and (c). positivity, i.e.,
all eigenvalues of $\rho(\textbf{u})$ are non-negative.

Indeed, the operator
$\frac{1}{N}(I+\sqrt{\frac{N(N-1)}{2}}\overrightarrow{\lambda}.u)$
automatically satisfies the conditions (a) and (b). However, not
every vector $\textbf{u}$, $|\textbf{u}|\leq 1$, allows
$\rho(\textbf{u})$ satisfies the positive  condition (c), for
example, see \cite{Chen054304,Ma064325}.

In the case of bipartite quantum systems
($H=\mathbb{C}^n\otimes\mathbb{C}^n$) composed of subsystems
$A$ and $B$, we can analogously represent the density operators
as
\begin{equation}\label{bipartitebloch}
\rho=\frac{1}{n^{2}}(I_{n}\otimes I_{n}+\sum\limits_{i=1}^{n^{2}-1}r_{i}\lambda_{i}\otimes
I_{n}+\sum\limits_{j=1}^{n^{2}-1}s_{j}I_{n}\otimes\tilde{\lambda}_{j}+\sum\limits_{i,j=1}^{n^{2}-1}t_{ij}\lambda_{i}\otimes\tilde{\lambda}_{j}),
\end{equation}
where $\lambda_i$  are the generators of
$SU(n)$. Notice that $\textbf{r}\in \mathbb{R}^{n^2-1}$
and $\textbf{s}\in \mathbb{R}^{n^2-1}$ are the coherence vectors of
the subsystems, so that they can be determined locally,
\begin{eqnarray}
\rho_A=\textrm{Tr}_B\rho&=&\frac{1}{n}(I_{n}+r_i\lambda_i),\nonumber\\
\quad\rho_B=\textrm{Tr}_A\rho&=&\frac{1}{n}(I_{n}+s_i\tilde{\lambda}_i).
\end{eqnarray}
The coefficients $t_{ij}$, responsible for the possible
correlations, form the real matrix $T\in \mathbb{R}^{(n^2-1)\times
(n^2-1)}$, and, as before, they can be easily obtained by
\begin{equation}\label{T}
t_{ij}=n^{2}\textrm{Tr}(\rho\lambda_i\otimes\tilde{\lambda}_j)=n^{2}\langle\lambda_i\otimes\tilde{\lambda}_j\rangle.
\end{equation}



Now consider observables represented by the following set of hermitian
operators:
\begin{eqnarray}
\hat{O}_{k}&=&\lambda_{i}^{a}\otimes\lambda_{j}^{b}, \nonumber\\
\hat{O}_{(n^{2}-1)^{2}+1}&=&\vec{z}\ldotp\vec{\lambda}^{a}\otimes\mathbf{I}^{b}+\mathbf{I}
^{a}\otimes\vec{w}\ldotp\vec{\lambda}^{b},
\end{eqnarray}
where $i,j=1,2,3...n^{2}-1$, for $i=j=1$, $k=1$, for $i=1, j=2$, $k=2$, and so on, so $k=1,2...(n^{2}-1)^{2}$. And $\vec{z},\vec{w}\in\Re^{n^{2}-1}$ with $||\vec{z}||=||\vec{w}
||=1$. We observe that the directions $\vec{z}$ and $\vec{w}$ can be picked out randomly.
Now we consider a relation among these observables as follows
\begin{equation}
W_{\rho}=\sum_{i< j}^{(n^{2}-1)^{2}+1}|\langle
\hat{O}_{i}\rangle_{\rho}\langle \hat{O}_{j}\rangle_{\rho}|,
\label{witness}
\end{equation}
where $\langle \hat{O}_{i}\rangle_{\rho}=Tr\{\hat{O}_{i}\rho\}$ and $|x|$ is
the absolute value of $x$. We see that $W_{\rho}=0$ if and only if the average
value of at least $(n^{2}-1)^{2}$ of the $(n^{2}-1)^{2}+1$ observables defined above is zero.

So, if $W_{\rho}=0$,  then $\rho$ must be one of the following form:
$$X_{ij}=\rho=\frac{1}{n^{2}}(I_{n}\otimes I_{n}+t_{ij}\lambda_{i}\otimes\tilde{\lambda}_{j}),$$
$$X_{(n^{2}-1)^{2}+1}=\frac{1}{n^{2}}(I_{n}\otimes I_{n}+\sum\limits_{i=1}^{n^{2}-1}r_{i}\lambda_{i}\otimes
I_{n}+\sum\limits_{j=1}^{n^{2}-1}s_{j}I_{n}\otimes\tilde{\lambda}_{j}),$$
where $i,j=1,2...n^{2}-1$.

The above $X_{i}$ are all classical states, i.e., they are the form
of Eq. (1).

\end{document}